\documentclass[fleqn,usenatbib]{mnras}

\usepackage{newtxtext,newtxmath}
\usepackage[T1]{fontenc}

\DeclareRobustCommand{\VAN}[3]{#2}
\let\VANthebibliography\thebibliography
\def\thebibliography{\DeclareRobustCommand{\VAN}[3]{##3}\VANthebibliography}

\usepackage{graphicx}	
\usepackage{amsmath}	
\usepackage{wasysym}           
\usepackage{graphicx}
\usepackage{epstopdf}
\usepackage{mathrsfs}
\usepackage{anyfontsize}
\usepackage{natbib}
\usepackage{color}
\usepackage{lipsum}
\usepackage{diagbox}

\title[Nonthermal Filaments]{Nonthermal Filaments from the Tidal Destruction of Clouds in the Galactic Center}

\author[Coughlin, Nixon, \& Ginsburg]{
Eric R.~Coughlin, $^{1}$\thanks{E-mail: ecoughli@syr.edu}
C.~J.~Nixon, $^{2}$
Adam Ginsburg$^{3}$
\\
$^{1}$Department of Physics, Syracuse University, Syracuse, NY 13244, USA \\
$^{2}$School of Physics and Astronomy, University of Leicester, Leicester, LE1 7RH, UK \\
$^{3}$Department of Astronomy, University of Florida, P.O. Box 112055, Gainesville, FL, USA
}

\date{Accepted XXX. Received YYY; in original form ZZZ}

\pubyear{2020}

\begin{document}
\label{firstpage}
\pagerange{\pageref{firstpage}--\pageref{lastpage}}
\maketitle

\begin{abstract}
Synchrotron-emitting, nonthermal filaments (NTFs) have been observed near the Galactic center for nearly four decades, yet their physical origin remains unclear. Here we investigate the possibility that NTFs are produced by the destruction of molecular clouds by the gravitational potential of the Galactic center. We show that this model predicts the formation of a filamentary structure with length on the order of tens to hundreds of pc, a highly ordered magnetic field along the axis of the filament, and conditions conducive to magnetic reconnection that result in particle acceleration. This model therefore yields the observed magnetic properties of NTFs and a population of relativistic electrons, without the need to appeal to a dipolar, $\sim$ mG, Galactic magnetic field. As the clouds can be both completely or partially disrupted, this model provides a means of establishing the connection between filamentary structures and molecular clouds that is observed in some, but not all, cases. 
\end{abstract}


\section{Introduction}
The Galactic center (GC) harbors an anomalously high gas density and temperature (e.g., \citealt{gusten80, gusten81, bally87, ferriere07}) and a bright radio and X-ray source that is generally accepted to be a supermassive black hole (SMBH; e.g., \citealt{balick74, backer99, baganoff03}) that has its own population of gravitationally bound stars (e.g., \citealt{ghez03, ghez08, gillessen09, habibi17}), at least one inclined disc of young stars (e.g., \citealt{genzel03, lu09}), and 30 or more Wolf-Rayet stars (e.g., \citealt{paumard01, martins07, ressler18}). These (and other) properties (see, e.g., \citealt{oort77, morris96, genzel10} for reviews) make the GC among the most dynamic and rich environments in the local Universe.

The GC contains a number of radio-bright filaments, which are extended ($\gtrsim$ tens of pc long) and narrow ($\lesssim 0.5$ pc wide) tendrils of gas, the observations and analyses of which (e.g., \citealt{ekers83, yusef84, morris85, yusef86, yusef87, bally89, gray91, uchida92, gray95, morris96, lang99, larosa00, bicknell01b, bicknell01, larosa04, nord04, yusef04, morris14, yusef16, morris17, heywood19, wang20, yusef20}) find that they possess the following properties:

\begin{enumerate}
\item{Their radio emission is nonthermal, and hence they are referred to as nonthermal filaments (NTFs); the emission is thought to arise from synchrotron-emitting electrons in the NTF magnetic field.}
\item{The magnetic field of each NTF is highly ordered and approximately parallel to its axis, as indicated by polarization studies (which also substantiate the notion that synchrotron is the dominant process contributing to the radio emission).}
\item{The longest and earliest-observed NTFs were found to be aligned nearly perpendicularly to the Galactic disc.}
\item{In contrast to the longest NTFs, relatively short filaments (also observed more recently) maintain more random orientations with respect to the Galactic disc.}
\item{NTFs are located preferentially within a few degrees of the GC and are therefore seemingly unique to the central region of the Galaxy.}
\item{A fraction of NTFs terminate at thermal sources that appear to be associated with star-forming regions or molecular clouds.}
\end{enumerate}
The combination of properties 2 and 3 led to the suggestion that NTFs trace the background, and hence approximately dipolar, magnetic field associated with the GC. However, more recent observations of shorter filaments and their less ordered distribution (property 4) suggest that, if NTFs do indeed trace the GC magnetic field, it may not be so well-ordered. \citet{larosa05} found that the diffuse, nonthermal emission originating from the GC implies a relatively weak field ($\sim$ tens of $\mu$G) compared to the large ($\sim$ mG) values inferred in filaments \citep{morris96}, the latter inference based upon the assumption that the magnetic field pressure within a filament must balance the turbulent ram pressure of the GC gas to maintain its linear morphology; more recent observations of molecular lines suggest Galactic magnetic field strengths on the order of $\sim 100$ $\mu$G \citep{oka19}. \citet{yusef05}, assuming inverse Compton scattering was producing X-rays in one NTF, inferred a magnetic field strength of $\sim 80\mu$G within that filament, while \citet{gray95} found from Faraday rotation that the immediate vicinity of the filament G359.1-00.2, or the ``Snake,'' is characterized by a $\sim 10\mu$G field. These latter measurements contrast the strong fields based on dynamical arguments.

In addition to the strength and origin of the filamentary magnetic fields, a separate (but not unrelated) puzzle concerns the origin of the filaments themselves. \citet{gray95} (see also \citealt{bicknell01b}) delineated a number of formation mechanisms for the Snake, including shock fronts, star wakes, and cosmic strings, and concluded that none is particularly well-suited for explaining all of its properties. Others have investigated the interaction between fast-moving clouds and a pre-existing magnetic field as a possible origin (e.g., \citealt{benford88,staguhn98}), and some have argued that filaments can be generated as a consequence of the collision between stellar winds in star-forming regions (e.g., \citealt{rosner96, yusef03}). More recently it has been suggested that an outflow, or wind, emanating from the GC could interact with giant molecular clouds and produce filamentary structures (e.g., \citealt{banda18, yusef19}). The radio observations of the GC by MeerKAT \citep{heywood19} indicate that some of the most extended filamentary structures (the ``Arc'') arise coincidently with the longitudinal extremities of inflated radio bubbles, and are therefore likely edge-brightened emission from that same, large-scale source (see Figure 2 of \citealt{heywood19}). 

While edge brightening of the radio bubbles appears to be a likely explanation for the Arc, the numerous other, smaller filaments -- at least dozens in number from Figure 2 of \citealt{heywood19} -- do not seem to be at least directly related to this phenomenon. Here we analyze a distinct model for the origin of these filaments, being the destruction of a molecular cloud by the tides of the Galactic potential (dominated by the SMBH at sufficiently small radii; see Section \ref{sec:model} below). This model -- in the limit that the SMBH dominates the potential -- was originally proposed by \citet{ekers83} and analyzed quantitatively and numerically by \citet{quinn85} and \citet{sanders98}, and has been more recently considered in the context of star formation in the GC \citep{bonnell08, alig11}. However, as we discuss in more detail below, these investigations were either in a regime conducive to the formation of a ring of gas \citep{quinn85, sanders98}, or were extreme in that the point of closest approach to the SMBH was comparable to the size of the cloud itself \citep{bonnell08, alig11}. We show that under different (and probably more likely) circumstances, the destruction of a gas cloud by a SMBH creates a nearly linear filament with a magnetic field oriented parallel to its axis. Moreover, while the prediction of this model is that the orientation of the field should be along the filament axis, the \emph{sign} of the field is not necessarily the same everywhere within the filament; this model therefore also predicts that there should be regions within the filament that contain nearly parallel lines of magnetic field but with opposing signs, which generate current sheets that are conducive to reconnection and particle acceleration. In addition to providing an ordered field, this formation mechanism therefore also establishes a means of nonthermal particle acceleration. 

In Section \ref{sec:model} we describe the basic dynamical model and the consequences of this model in the context of explaining the properties of filaments (points 1 -- 6 above). We discuss and conclude in Section \ref{sec:summary}.

\section{Dynamical Model}
\label{sec:model}
Gaseous clouds near the GC can, in principle, have a range of sizes, from giant molecular clouds with radii on the order of tens of pc and masses $10^{4}$ -- $10^{7} M_{\odot}$ (e.g., \citealt{blitz93, fukui10}), to objects such as G2 and G1 (e.g., \citealt{gillessen12, burkert12, gillessen13, pfuhl15}) that possess masses comparable to that of the Earth. Such a cloud can exist relatively unperturbed, i.e., can retain approximate spherical symmetry\footnote{We acknowledge that realistic molecular clouds are by no means perfectly spherical, and possess a range of aspherical deformations that are likely indicative of their violent formation process; e.g., \citet{dobbs13}. For the purpose of this theoretical investigation, however, we maintain this assumption. }, until it nears a distance to the GC where tides from the Galactic potential overwhelm its self-gravity. If the SMBH were the only contributor to the gravitational field at $\sim$ pc scales, then this distance -- the tidal radius $r_{\rm t}$ -- is given by the location at which the differential gravitational force of the SMBH across the cloud diameter equals the self-gravity of the cloud, or\footnote{This expression assumes that the cloud is bound by self-gravity initially; if this is not the case and the cloud is a transient feature (e.g., \citealt{dobbs11}), then the effective tidal radius of the over-density can be much larger.} (e.g., \citealt{hills75})

\begin{multline}
\frac{GM_{\bullet}R_{\rm c}}{r_{\rm t}^3} \simeq \frac{GM_{\rm c}}{R_{\rm c}^2} \Rightarrow r_{\rm t} \simeq R_{\rm c}\left(\frac{M_{\bullet}}{M_{\rm c}}\right)^{1/3} \simeq \left(\frac{M_{\bullet}}{\rho_{\rm c}}\right)^{1/3} \\ \simeq 30\textrm{ pc} \left(\frac{M_{\bullet}}{4.15\times10^{6}M_{\odot}}\right)^{1/3}\left(\frac{\rho_{\rm c}}{10^{-20}\textrm{g cm}^{-3}}\right)^{-1/3}. \label{rtidal}
\end{multline}
Here the cloud has a radius $R_{\rm c}$, a mass $M_{\rm c}$, and a density\footnote{The density $\rho_{\rm c}$ here is the average cloud density, though more realistically we expect turbulent velocities to create over- and under-dense regions within the cloud that can be varied around this value. We return to the implications of this in Section \ref{sec:summary} below.} $\rho_{\rm c}$, and the SMBH has a mass $M_{\bullet}$; in the last line we scaled $M_{\bullet}$ and $\rho_{\rm c}$ to the fiducial values $M_{\bullet} = 4.15\times10^{6}M_{\odot}$ \citep{gravity19} and $\rho_{\rm c} = 10^{-20}$ g cm$^{-3}$ (characteristic of a number density of $\sim 10^{4}$ cm$^{-3}$). 

In addition to the SMBH, the GC has a number of distinct, stellar components that contribute to the mass profile contained within spherical radius $r$, which we denote $M(r)$. With a mass of $M_{\bullet} \simeq 4\times 10^{6}\,M_{\odot}$, from Figure 14 of \citet{launhardt02} the mass contained within the nuclear stellar cluster equals that of the SMBH at a radius of $\sim 3$ pc, and the stellar cluster then dominates the potential out to a radius of $\sim 20-30$ pc. Outside of $\sim 30$ pc, the nuclear stellar disc contributes predominantly to the mass profile. As a consequence, for radii $r \lesssim 30$ pc for which the stellar cluster and SMBH dominate, the enclosed mass profile displays a fairly shallow increase in radius, and with $M(r) \propto r^{\alpha}$ has $\alpha \lesssim 1$. However, as described in Appendix A of \citet{kruijssen15}, the dominance of the disc at larger radii implies that the mass profile steepens to $\alpha \simeq 2$, with $\alpha$ best fit by $\alpha = 2.2$ over the range $40 \lesssim r \lesssim 100$ pc. More recently, \citet{sormani20} re-evaluated the mass profiles within the central $\sim 100$ pc of the GC by using line of sight velocity kinematics of the gas (whereas the estimates in \citealt{launhardt02} were based on photometry), and concluded that the amount of mass contained in the nuclear stellar disc is likely on the low end of the estimates from \citet{launhardt02} and is more highly concentrated. As such, their best-fit mass profiles remain shallower in their slope than those inferred by \citet{launhardt02}, and never quite steepen to the point where $\alpha > 2$.

Importantly, and as also emphasized in \citet{dale19} and \citet{kruijssen19}, a steep increase in $M(r)$ as a function of radius implies that the tidal field can become \emph{compressive} in the radial direction, and instead of destroying clouds of gas can actually result in their further contraction. Specifically, since the gravitational field is just\footnote{This strictly only applies when the gravitational field is spherically symmetric, but this approximation is likely upheld reasonably accurately near the GC where stars are highly concentrated \citep{dale19} and is appropriate for this level of discussion.} $\propto M(r)/r^2$, any region that has $\alpha > 2$ produces a stronger (weaker) gravitational field at large (small) radii and results in the radial compression of a cloud with finite radius. Consequently, even though Equation \eqref{rtidal} suggests that clouds with sufficiently low density can be disrupted at fairly large radii by the gravitational potential of the SMBH, the simultaneous dominance of the disc mass and the steep value of $\alpha$ for radii $\gtrsim 60$ pc (see Figure A1 of \citealt{kruijssen15}) implies that this is likely not the case; instead, gas overdensities may be relatively unperturbed by the tidal potential in this region and their formation even \emph{promoted} by tides \citep{kruijssen19}. For the shallower increase in $M(r)$ as found from \citet{sormani20}, the tidal potential is never completely compressive, but the steeper increase in $M(r)$ from the contribution of the nuclear stellar disc still implies that the GC tides are relatively weak at distances of $\sim 100$ pc compared to those at smaller distances.

Our model will assume that a cloud is deposited onto a low angular momentum orbit about the SMBH (and nuclear stellar cluster) that takes it within the distance $\sim r_{\rm t}$; while the formation mechanism of such a cloud is uncertain, the steep increase in the mass profile at larger radii and the simulations of \citet{dale19} and \citet{ kruijssen19} suggest that it is at least plausible that they can be formed -- if somewhat violently and stochastically \citep{dobbs13} -- at distances somewhat larger than $r_{\rm t}$. Typical cloud lifetimes of $\sim 10$ Myr \citep{dobbs13} are comparable to the freefall time from 100 pc if the total enclosed mass is $\sim 10^{8} M_{\odot}$, meaning that even without the compressive nature of the tidal field, clouds should be sufficiently long-lived to travel from their formation site to the tidal radius. It is also possible that such clouds form much farther out, and approach the tidal radius on near-parabolic orbits (e.g., \citealt{ridley17, sormani18, tress20, sormani20b}). After they form, collisions (e.g., \citealt{dobbs15}) and additional tidal interactions may dissipate energy to the point where they reach pericenter distances within $\sim 50$ pc (where the density profile is sufficiently concentrated that the tidal field becomes disruptive); alternatively, gravitational scattering may abruptly place a cloud onto a low angular momentum \emph{and} low binding energy orbit, causing it to plunge into the central $\lesssim 50$ pc region. 

For cloud radii that satisfy $R_{\rm c} \lesssim few$ pc, Equation \eqref{rtidal} demonstrates that the \emph{tidal approximation}, which maintains only leading-order terms in the ratio of the cloud radius to the position of the center of mass when considering the dynamics of the cloud within the tidal sphere (e.g., \citealt{sari10, stone13}), is upheld until the pericenter distance of the COM, $r_{\rm p}$, becomes comparable to the radial extent of the cloud itself. Defining the ratio $\beta \equiv r_{\rm t}/r_{\rm p}$, such that a larger $\beta$ implies a deeper, and more disruptive, encounter, the breakdown of this approximation occurs when $r_{\rm p} \sim R_{\rm c}$. For cloud radii on the order of $\sim 10$ pc, this approximation is only valid for $\beta \sim few$, but for smaller clouds can be upheld for $\beta$ in excess of 10. However, it is worth noting that the spread in energies of the gas parcels comprising the cloud -- which determines the rate at which the cloud is deformed by the tidal field (see Sections \ref{sec:energy} and \ref{sec:spread}) -- depends only on the properties of the cloud at the tidal radius provided that the orbits of gas parcels remain ballistic within the tidal sphere. Thus, if the tidal approximation is upheld at the tidal radius, then the spread in energies is well constrained independently of the pericenter distance.

A simple approximation for understanding the evolution of the cloud as it passes within $r_{\rm t}$ can be obtained by treating the interaction between the cloud and the object that creates the disruptive potential as occurring impulsively, wherein the cloud retains hydrostatic balance prior to reaching $r_{\rm t}$ and is completely dominated by the gravitational field of the central object -- such that gas parcels comprising the cloud follow ballistic orbits -- within $r_{\rm t}$ \citep{lacy82, lodato09, stone13, coughlin19}. Here the central object is the combination of the nuclear stellar cluster and the SMBH; the former dominates the mass at relatively large radii, but the shallower slope of $M(r)$ owing to this component implies that the tidal influence of the latter can still be dynamically important. For the sake of simplicity and concreteness, we consider only the case where the SMBH contributes to the gravitational field, which is more accurate for smaller pericenter distances ($\beta > 1$) and cloud radii. However, the qualitative results should still apply for more general potentials, provided that the potential is sufficiently concentrated and hence the mass profile is relatively shallow with radius (i.e., has $M(r) \lesssim r^{2}$). 

We do not analyze the details of individual orbits of gas parcels under the impulse approximation, which has already been done extensively in the context of tidal disruption events (TDEs; e.g., \citealt{stone13, coughlin20}); rather, we elucidate the general consequences of the tidal and impulse approximations on the cloud dynamics in relation to the observed properties of NTFs. 

\subsection{Filament formation}
\label{sec:filament}
Because the orbits of gas parcels are ballistic in the (effectively) Newtonian potential of the SMBH in this model, the orbital timescale of any given gas parcel is governed solely by its initial distance from the SMBH and the energy and angular momentum of the cloud COM\footnote{This is also true for the more generic case in which the gravitational potential is solely a function of radius and hence when one accounts for the additional impact of the nuclear stellar cluster.}. Thus, gas parcels at the same initial distance from the SMBH remain relatively close to one another compared to those at different distances, and the gas cloud is therefore stretched preferentially in a single direction. The transformation of the cloud into a filamentary structure is thus unavoidable as a consequence of the tidal interaction alone, and this feature can only be violated -- with the interaction much more violent and less ordered -- if the pericenter distance of the cloud is comparable to or less than the radial extent of the cloud itself. The latter scenario is the one considered by, e.g., \citet{bonnell08, alig11}. 

\begin{figure*}
   \centering
   \includegraphics[width=0.995\textwidth]{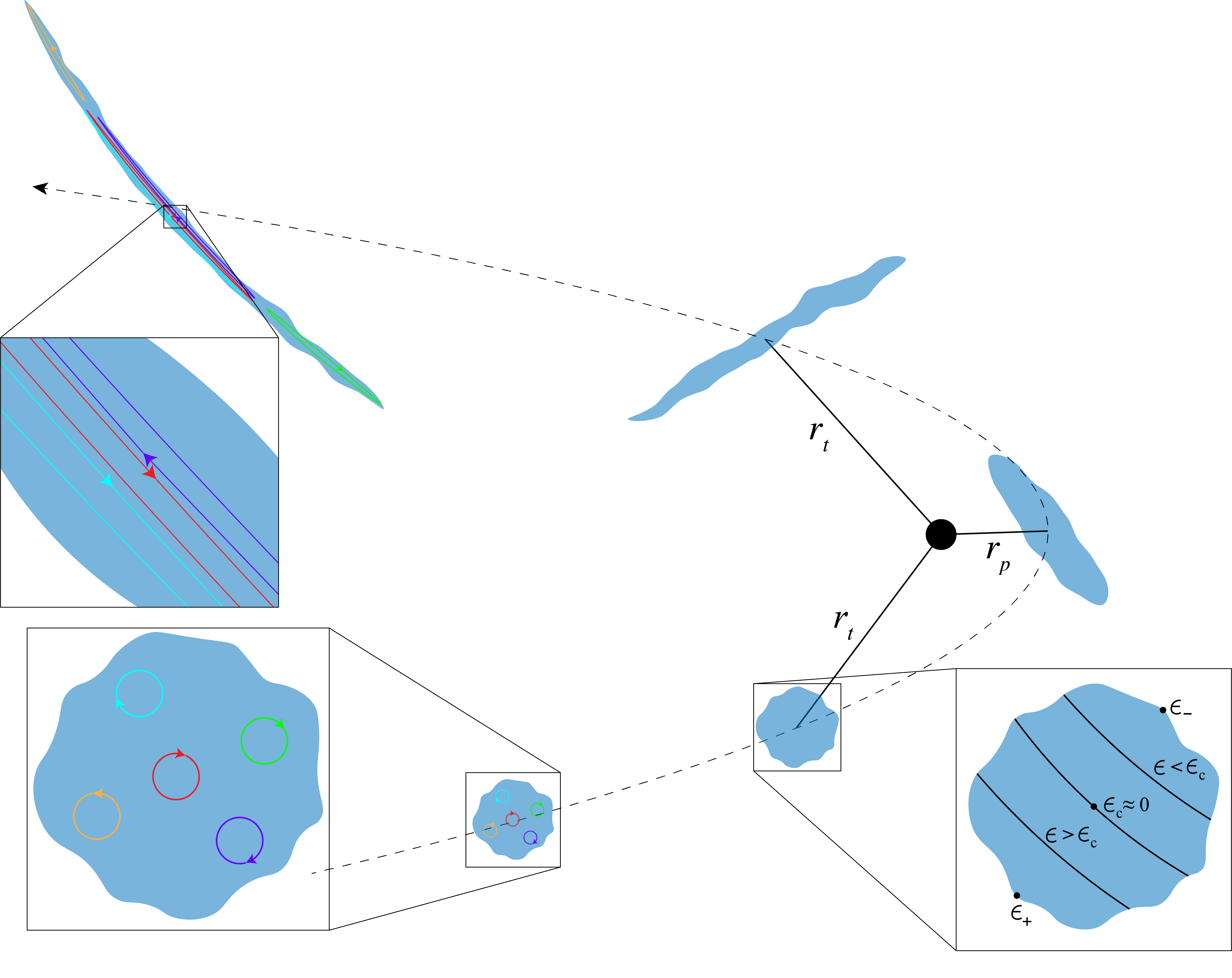} 
   \caption{A qualitative illustration of the destruction of a gas cloud by the gravitational potential of the GC; the dashed line shows the orbit of the center of mass, which passes through the tidal radius $r_{\rm t}$ and has a pericenter distance to the GC of $r_{\rm p} \le r_{\rm t}$. The cloud remains relatively unperturbed by the tidal distortion of the gravitational potential and is approximately spherical until it reaches the tidal radius, $r_{\rm t}$. At the tidal radius, the differential gravitational force across the cloud diameter roughly equals the self-gravity of the cloud, leading to its destruction. Interior to the tidal radius, the evolution of gas parcels within the cloud can be approximated as ballistic in the gravitational field of the GC, which results in the stretching of the cloud preferentially in the radial direction and its transformation into a filamentary structure. The inset on the right, which coincides with when the cloud enters the tidal radius, shows the location of the most bound ($\epsilon_{-}$) and most unbound ($\epsilon_+$) gas parcels, as well as the energy of the center of mass $\epsilon_{\rm c}$. Since the entire cloud is assumed to move with the center of mass upon entering the tidal disruption radius, surfaces of constant specific energy coincide with surfaces of constant distance from the GC, which are shown by the approximately linear curves in this inset. The colored loops, also shown in the insets on the left, indicate examples of closed magnetic field lines within the cloud, and show that -- under the assumption that magnetohydrodynamics applies -- the frozen-in nature of the field lines results in their preferential alignment with the axis of the filament. The random orientation of field lines within the initial cloud implies that the current sheet that forms as a consequence of this stretching is conducive to magnetic reconnection, which can energize electrons and give rise to the synchrotron emission that we observe from NTFs. }
   \label{fig:cloud_disruption}
\end{figure*}

Figure \ref{fig:cloud_disruption} provides a visualization of this disruption and filament formation process: the cloud is approximately unperturbed (and assumed to be roughly spherical) until it reaches the tidal radius, $r_{\rm t}$, given in Equation \eqref{rtidal} for fiducial parameters. After this point, gas parcels within the cloud follow ballistic orbits, which results in the stretching of the cloud in the direction of the GC. Because the potential only depends on spherical radius (to our level of approximation), gas parcels at the same initial distance from the SMBH when the cloud passes through the tidal radius remain relatively close to one another, resulting in the formation of a filament. As we describe in more detail in Section \ref{sec:caustics} below, the impulse and tidal approximations result in the compression of the gas in the transverse (non-radial) directions, which serves to keep the filament thin after it passes through pericenter ($r_{\rm p}$ in this figure). The dashed line shows the orbit of the cloud COM, assumed to be roughly parabolic in this case (see Section \ref{sec:energy} below). The colored lines give examples of magnetic field lines within the cloud, which are stretched along with the gas if magnetohydrodynamics applies (see Section \ref{sec:magnetic} for further discussion).

\subsection{Energy dependence}
\label{sec:energy}
At the time the cloud enters the tidal radius, the application of the impulse approximation shows that the orbital energy of a given gas parcel is

\begin{equation}
\epsilon = \frac{1}{2}v^2-\frac{GM_{\bullet}}{r} \simeq \epsilon_{\rm c}+\frac{GM_{\bullet} R_{\rm c}}{r_{\rm t}^2}\eta,
\end{equation}
where in the last line we defined the distance of the gas parcel to the black hole $r$ through $r = r_{\rm c}+\eta R_{\rm c}$, so $-1 \le \eta \le 1$, and $\epsilon_{\rm c}$ is the energy of the center of mass. We also adopted the tidal approximation, and therefore dropped terms that scale as $\propto \left(R_{\rm c}/r_{\rm t}\right)^2$ and higher in this expression. This shows that the energies of the most unbound gas parcel (with $\eta = 1$) and most tightly bound gas parcel (with $\eta = -1$) are

\begin{equation}
\epsilon_{\pm} = \frac{GM_{\bullet}}{r_{\rm p}}\left(\frac{1}{2}\left(e-1\right)\pm\frac{1}{\beta}\left(\frac{M_{\rm c}}{M_{\bullet}}\right)^{1/3}\right), \label{epm}
\end{equation}
where the plus (minus) sign is for the most unbound (bound) gas parcel, and that surfaces of constant energy within the cloud coincide with surfaces of constant distance from the GC. These surfaces and the locations of the most bound and unbound gas parcels within the gas cloud at the time it enters the tidal disruption radius are shown qualitatively in Figure \ref{fig:cloud_disruption}. We also used the expression for the energy of the cloud COM in terms of its eccentricity, $e$, and pericenter distance, $r_{\rm p}$. The first term in parentheses represents the energy of the center of mass, while the second term is an additional energy spread (as a function of location within the cloud) that arises from the tidal potential of the SMBH. Equation \eqref{epm} demonstrates that there is a dichotomy in the behavior of the cloud as a function of the eccentricity of the center of mass: for tightly bound\footnote{By referring to the cloud as bound we mean gravitationally with respect to the SMBH, not to itself.} clouds on more circular orbits, $e \ll 1$, and the positive energy imparted to the most unbound gas parcel owing to the tidal field (the second term in parentheses in Equation \ref{epm}) is not sufficient to completely unbind it from the SMBH. In this case, the cloud is transformed into a roughly circular filament that wraps around the black hole and generates an eccentric disc; this situation is the one analyzed by \citet{quinn85} and \citet{sanders98} (see also \citealt{nixon20}). 

There is, however, another possibility, which is that the COM of the cloud is only very weakly bound to the SMBH ($e \simeq 1$) and the second term exceeds the first, producing unbound material even if the cloud were initially (weakly) bound; this situation is achieved naturally if the cloud is formed at much larger distances than the tidal radius, where the only way to reach the GC is to be placed on a low-angular momentum orbit that -- by virtue of the fact that the initial distance is much greater than $r_{\rm t}$ -- has a very low binding energy. Instead of producing a tightly bound ring of eccentric gas, the disruption of marginally bound clouds results in the ejection of roughly half of the cloud mass, while the other half is more tightly bound (than the original cloud COM) and returns to the SMBH. The return time of the most bound debris in the case of a marginally bound ($e = 1$) orbit of the COM, which we denote $T_{\rm ret}$, is then

\begin{multline}
T_{\rm ret} = 2\pi\left(\frac{R_{\rm c}}{2}\right)^{3/2}\frac{M_{\bullet}}{M_{\rm c}\sqrt{G M_{\bullet}}} \\
\simeq 26\textrm{ Myr}\left(\frac{R_{\rm c}}{2.5 \textrm{ pc}}\right)^{3/2}\left(\frac{M_{\bullet}}{4\times10^{6}M_{\odot}}\right)^{1/2}\left(\frac{M_{\rm c}}{10^4M_{\odot}}\right)^{-1}, \label{Tret}
\end{multline}
where in the last line we scaled the result to fiducial values (the cloud radius of $2.5$ pc ensures that the average density of the cloud is $\simeq 10^{-20}$ g cm$^{-3}$ given a cloud mass of $10^{4}M_{\odot}$). This expression can be derived by using the energy of the most bound debris (the negative solution in Equation \ref{epm} with $e =1$) and the relationship between the energy of a Keplerian orbit and its orbital time, $T \propto (-\epsilon)^{-3/2}$, and making a few algebraic simplifications (e.g., \citealt{rees88}). The center of mass of the cloud approximately follows a radial orbit with $r_{\rm c} \propto t^{2/3}$, and hence in this time the radial position of the cloud COM recedes to a distance

\begin{equation}
r_{\rm c} \simeq \left(\frac{3}{2}\sqrt{2GM_{\bullet}}T_{\rm ret}\right)^{3/2} \simeq r_{\rm t}\left(\frac{M_{\bullet}}{M_{\rm c}}\right)^{1/3},
\end{equation}
which is generally a few to ten times the tidal radius of the cloud for typical mass ratios of the mass of the SMBH to that of a given molecular cloud, or tens to hundreds of pc. 

\subsection{Filament spreading}
\label{sec:spread}
Even though roughly half of a marginally bound cloud will return to the SMBH on the timescale given by Equation \eqref{Tret}, initially the orbits of gas parcels within the cloud are all narrowly confined about $r_{\rm c}$, within which the tidal approximation is valid. The rate at which individual orbits diverge over time as a consequence of the energy spread imparted by the tidal potential can be estimated in a number of ways (e.g., \citealt{coughlin16}); one straightforward method of doing so is to recall the energy integral for a given fluid element,

\begin{equation}
\frac{1}{2}\dot{r}^2+\frac{1}{2}\frac{\ell^2}{r^2}-\frac{GM_{\bullet}}{r} = \epsilon,
\end{equation}
write $r \simeq r_0(t)+\epsilon r_1(t)+\ell^2r_2(t)+\mathcal{O}[\epsilon^2,\ell^4,\epsilon\ell^2]$, and equate the leading-order terms in the specific energy, $\epsilon$, and the square of the specific angular momentum, $\ell^2$ (in this expansion $r_0$ is the marginally bound orbit with $\epsilon = \ell^2 = 0$). When the COM is receding from the SMBH, i.e., post-pericenter, this method yields

\begin{equation}
r_1 \propto r_0^2 \propto t^{4/3}. \label{r1oft}
\end{equation}
This scaling describes the leading-order (in energy) spreading of material from the marginally bound ($\epsilon = 0$) orbit and becomes more accurate as the material recedes beyond the tidal radius, while a more detailed analysis of the orbital motion is needed to understand the evolution within $r_{\rm t}$. However, if we use the fact that the material is generally stretched to a length of $\sim few\times R_{\rm c}$ when the COM reaches $r_{\rm t}$ (see, e.g., \citealt{coughlin20}), then by the time the COM recedes to a distance of $\sim few \times r_{\rm t}$, Equation \eqref{r1oft} illustrates that the length of the filament has expanded to $\sim 10-100 R_{\rm c}$. Figure \ref{fig:cloud_disruption} gives an illustration of this evolution, and shows that the difference in energy across the cloud diameter at the tidal radius gives rise to an elongation of the filament.

Equation \eqref{r1oft} describes the radial spreading of the filament until $|\epsilon r_1| \simeq r_0$; after this time, which occurs roughly on the timescale $T_{\rm ret}$ for the most unbound and bound gas parcels, the behavior of the filament bifurcates around the $\epsilon = 0$ orbit. As described in \citet{coughlin16} in the context of TDEs, the radial spreading of the bound segment becomes more dramatic as fluid elements start to return to the SMBH, while that of the unbound region is less pronounced and -- because unbound elements of the filament eventually approach constant velocities -- asymptotically spread as $\propto t$. We return to the implications of this behavior in the context of the filament density in Section \ref{sec:width} below.

\subsection{Orthogonal compression}
\label{sec:caustics}
The analysis in Section \ref{sec:spread} demonstrates that the filament spreads radially outside of $\sim r_{\rm t}$, but as described in Section \ref{sec:filament}, gas parcels at the same initial distance from the black hole within the cloud stay relatively close to one another. As the COM of the cloud approaches the SMBH (i.e., prior to reaching pericenter), the gas parcels out of the plane are compressed due to the tidal acceleration. If ballistic motion were upheld indefinitely, the top and bottom of the cloud would form a caustic as orbits cross the orbital plane of the cloud \citep{carter83, stone13}. Any finite pressure prevents the formation of a true caustic, with the pressure increase occurring adiabatically or through the formation of a shock \citep{bicknell83} as the cloud is compressed vertically. Simulations in the context of TDEs have found that relatively large $\beta$ are required\footnote{It is important to note that for TDEs the dominant contributor to the pressure arises from gas and radiation and hence the adiabatic index satisfies $4/3 \lesssim \gamma \lesssim 5/3$. If the cloud is nearly isothermal and radiative, such that $\gamma \simeq 1$, then the formation of a shock could occur at more modest $\beta$.} ($\beta \gtrsim 7$; \citealt{brassart08}) to generate a shock that propagates inward to prevent the collapse, and hence for most disruptions the pressure that builds to resist the vertical compression of the cloud proceeds adiabatically. 

More recently \citep{coughlin16b, coughlin20} it has been shown that the motion of the gas within the orbital plane is also compressive under the impulse approximation, which forms a distinct caustic within the plane as the COM of the cloud reaches $\sim r_{\rm t}$ on its way out from pericenter. This secondary caustic exists for deep encounters, and keeps the filament narrow within the plane; this is shown qualitatively in Figure \ref{fig:cloud_disruption}, where the width of the filament at the time it reaches $r_{\rm t}$ on its way out is much smaller than its length as a consequence of this in-plane compression. The rate at which the cloud compresses is also at most only mildly supersonic, implying that the pressure that builds to resist the in-plane convergence of the flow occurs adiabatically.  

This in- and out-of-plane compression that occurs as the cloud orbits within $\sim r_{\rm t}$ keeps the gas narrowly confined in the transverse (i.e., non-radial) directions, and results in a less dramatic decline in the density as compared to what would occur if the gas expanded more isotropically. If the cloud was in hydrostatic balance initially, then the nearly adiabatic compression within the plane augments the density of the filament to the point where it becomes self-gravitating in the transverse directions. The width of the filament is then initially determined by the balance between pressure and self-gravity, meaning that the width is comparable to the radial size of the cloud. 

\subsection{Filament pressure, width and density}
\label{sec:width}
There are at least three main sources of pressure within the initial (i.e., pre-disruption) molecular cloud, being thermal gas pressure, turbulent ram pressure, and magnetic pressure. For ordinary molecular clouds with gas temperatures on the order of $\sim 10$ K, the thermal pressure is sub-dominant to the turbulent pressure, as the former yields a thermal velocity of $\sim few\times 0.1$ km s$^{-1}$ while the turbulent velocities in molecular clouds are observed to be $\sim few$ km s$^{-1}$ (e.g., \citealt{larson81}). However, the temperatures in the GC are systematically higher than average with characteristic gas temperatures in the range $\sim 70$ -- $200$ K (e.g., \citealt{morris96}), which implies that thermal support could be more important in these circumstances, though the turbulent velocities are also typically larger by a comparable factor (perhaps hinting at a relation between the two; e.g., \citealt{wilson82}). The importance of magnetic pressure is less certain, though it is likely comparable to or somewhat less than that from turbulent pressure (e.g., \citealt{crutcher12, nixon19}; turbulent reconnection could further dissipate magnetic energy and reduce its relative importance; \citealt{lazarian05, santos10}). 

While the turbulent pressure within the cloud is likely dominant initially, as the cloud is compressed by the tidal field on its way to pericenter, the thermal component increases adiabatically to the point where it may become comparable to the turbulent pressure; the compression will also promote the merger of turbulent patches of gas, which serves to reduce the turbulent pressure. Furthermore, as the cloud is stretched and continues to elongate post-pericenter, the relative velocities along the direction of the filament (i.e., the shear induced by the tidal stretching) eventually become on the order of the turbulent velocity, which further inhibits the role of turbulent heating. The details of the magnetic field are considered in Section \ref{sec:magnetic} below, but if the field is initially highly disordered and the gas is sufficiently ionized that magnetohydrodynamics applies, the field strength initially scales with the density as $\propto \rho^{2/3}$ and the pressure is $\propto B^2 \propto \rho^{4/3}$ (e.g., \citealt{mestel66, mckee03}), which is just an adiabatic equation of state with $\gamma = 4/3$. 

With these points in mind, the long-term evolution of the filament width $H$, i.e., after the cloud COM recedes to distances $\gtrsim r_{\rm t}$, depends on the equation of state of the gas and how rapidly the filament is spreading radially. As we showed above, for small orbital energies and for times $\lesssim T_{\rm mb}$, the length of the filament is characterized by $L \propto t^{4/3}$, while at late times the length grows more (less) steeply with time for bound (unbound) segments of the filament; for the unbound stream in a point-mass potential, the length scales as $L \propto t$. If the filament is in hydrostatic balance in the transverse direction, which the orthogonal compression in Section \ref{sec:caustics} suggests is likely to be the case initially, then the equation of hydrostatic equilibrium in the cylindrical radial direction gives \citep{kochanek94, coughlin16}

\begin{equation}
\frac{p}{\rho} \simeq 2\pi G\rho H^2 \quad \Rightarrow \quad p\propto \frac{1}{L^2H^2},
\end{equation}
where $p$ is the pressure and in the last line we used the fact that the density of a Lagrangian fluid element $\rho$ is just inversely proportional to the volume, or $\rho \propto 1/(LH^2)$. Given the uncertainties associated with the importance of various components of the pressure and in the absence of a more detailed model, we consider the case where the gas is parameterized by an effective adiabatic index $\gamma$ such that $p \propto \rho^{\gamma} \propto (LH^2)^{-\gamma}$, in which case the rearrangement of the above expression shows

\begin{equation}
H \propto L^{\frac{2-\gamma}{2\left(\gamma-1\right)}}.
\end{equation}
This scaling yields $H \propto t^{\frac{2\left(2-\gamma\right)}{3\left(\gamma-1\right)}}$ for $L \propto t^{4/3}$ and $H \propto t^{\frac{2-\gamma}{2\left(\gamma-1\right)}}$ for $L \propto t$, and illustrates that the density of the marginally bound segment of the stream scales as $\rho \propto t^{-\frac{4}{3\left(\gamma-1\right)}}$, whereas the unbound portion scales as the \emph{shallower power-law} $\rho \propto t^{-\frac{1}{\gamma-1}}$. These scalings also demonstrate that a necessary condition for the gas to remain self-gravitating in the transverse direction is that $\gamma \ge 5/3$ for the marginally bound segment of the stream, while $\gamma \ge 4/3$ for the unbound segment, the reason being that the tidal field of the black hole scales as $\propto H/r^3$ (and hence dominates the self-gravity of the fluid, $\propto \rho H$, if the density falls off more steeply than $\propto r^{-3}$; here $r$ is the Lagrangian radial position of the fluid element). The former condition could be satisfied if radiative cooling is efficient and can keep the filament thin. For concreteness, if the primary constituent to the pressure is from a monatomic gas (which is not unreasonable at early times given the anomalously high gas temperatures in the GC and the in- and out-of-plane compression that arises from the caustics; see Section \ref{sec:caustics}), then $\gamma = 5/3$ and the marginally bound segment density scales as $\rho \propto t^{-2}$, while the unbound segment declines as $\rho \propto t^{-3/2}$. On the other hand, with $\gamma = 4/3$ -- appropriate to the case where the magnetic field is highly tangled initially and responsible for the pressure support -- then $\rho \propto t^{-4}$ in the marginally bound segment, while $\rho \propto t^{-3}$ in the unbound segment.

\subsection{Magnetic field}
\label{sec:magnetic}
The evolution of a magnetic field as a star is destroyed by a SMBH was investigated by \citet{guillochon17} and \citet{bonnerot17}; for the case of typical (i.e., low-mass, main sequence) stars, the presence of the magnetic field is dynamically irrelevant -- even to the transverse structure of the stream of stellar material -- until extremely late times for TDEs. Here, however, the fact that a large contributor to the initial cloud pressure could be in the form of magnetic fields implies that magnetic effects could be important sooner. 

If the magnetic field is frozen in to the gas and ideal magnetohydrodynamics applies, then the magnetic flux through any Lagrangian fluid element is conserved. If we denote the parallel (to the filament axis) component of the magnetic field by $B_{||}$ and the perpendicular component by $B_{\perp}$, then flux conservation implies that

\begin{equation}
B_{||} \propto \frac{1}{H^2}, \quad B_{\perp} \propto \frac{1}{LH}, \label{Bscalings}
\end{equation} 
and hence the ratio of the perpendicular to the parallel component of the magnetic field scales as

\begin{equation}
\frac{B_{\perp}}{B_{||}} \propto \frac{H}{L}.
\end{equation}
Thus, the perpendicular component of the field declines relative to the parallel component provided that $H$ scales more weakly with time than $L$; for the specific case where $\gamma = 5/3$, $H \propto t^{1/4}$ and $L \propto t$ for the unbound segment of the filament, and this ratio scales as $B_{\perp}/B_{||} \propto t^{-3/4}$. The magnetic field therefore becomes \emph{preferentially aligned with the filament axis}. This behavior is shown qualitatively in Figure \ref{fig:cloud_disruption}, where example field loops are shown as the colored curves in the first and last snapshots of the cloud and the insets on the left.

At late times, therefore, we can neglect the perpendicular component of the magnetic field, and the magnetic pressure $P_{\rm B}$ scales as $P_{\rm B} \propto B_{||}^2 \propto H^{-4}$; in the limit that gas pressure is responsible for establishing the balance between pressure and self-gravity, the magnetic pressure declines as $P_{\rm B} \propto t^{-1}$. This decline in the magnetic pressure is much weaker than the decline in the gas pressure, however, which scales as $P_{\rm gas} \propto \rho^{5/3} \propto t^{-5/2}$. As also argued in \citet{guillochon17} and \citet{bonnerot17} in the context of TDEs, this difference in scaling implies that filamentary structures generated through this process eventually become magnetically dominated. Once magnetic pressure overtakes gas pressure, the filament is likely no longer able to remain self-gravitating, and correspondingly the density drops much more rapidly as magnetic pressure pushes the structure apart at a faster rate. 

Additionally, while this analysis demonstrates that the field becomes preferentially aligned with the long axis of the filament, the sign of the magnetic field is not necessarily the same everywhere. In the purely illustrative limit that the field of the disrupted cloud is dipolar with the dipole axis either aligned or anti-aligned with the orbital angular momentum of the initial cloud, the preferential stretching of the field in the direction of the filament axis creates opposing field lines as we approach the filament axis from out of the plane. A current sheet is therefore generated along the filament axis where the curl of the magnetic field is maximized -- analogous to the magnetotail generated from the interaction between Earth's magnetic field and the solar wind -- that creates conditions that are conducive to magnetic reconnection. Such reconnection and the corresponding dissipation of magnetic energy can then energize electrons that are accelerated along the axis of the filament, gyrate around the field lines, and give rise to nonthermal synchrotron emission.

Of course, a more realistic molecular cloud magnetic field -- likely amplified and sustained by turbulence -- will not have the highly ordered structure that yields the single, large current sheet just described. However, in general any closed magnetic field loops within the cloud that are not perfectly orthogonal to the direction in which the cloud is stretched will necessarily be transformed into a current-sheet-like configuration; this is also shown qualitatively in the top inset of Figure \ref{fig:cloud_disruption}, where opposing magnetic field lines come into close contact as a consequence of the initial distribution of magnetic field loops. It therefore seems unavoidable that generic magnetic field topologies give rise to, if not globally, local regions along the filament axis that undergo magnetic reconnection, energize particles, and those particles then gyrate around the field lines to yield synchrotron emission.

\subsection{Filament lifetime}
\label{sec:lifetime}
Section \ref{sec:width} illustrates that the density within the filament declines with time in a way that depends on energy; for tightly bound segments of the filament, the density initially declines in a manner comparable to the marginally bound segment, but eventually (i.e., on the timescale $T_{\rm ret}$ given in Equation \ref{Tret}) even more steeply as the radial shear along the filament increases as material accelerates toward the SMBH. Over a modest fraction of the return time of the most bound portion of the filament, the density of the tightly bound debris likely declines to the point where it is comparable to that of the ambient density, at which time it becomes susceptible to Kelvin-Helmholtz instabilities -- further amplified by the fact that the bound segment of the filament accelerates toward the SMBH -- that result in its disintegration into the background gas (see \citealt{burkert12} for a more detailed analysis in the context of the cloud G2). It is therefore likely that the bound segment of the filament does not survive for much longer than $T_{\rm ret}$ given in Equation \eqref{Tret}.

On the other hand, the density of the unbound material declines less steeply with time, and in a way that depends on the cross-sectional width of the filament (which itself depends on the contribution from magnetic pressure and gas pressure). In general, we expect the density of the unbound segment to scale as $\rho \propto t^{-1-\beta}$, where $\beta \ge 0$ encodes the contribution from the spreading of the filament in the transverse direction. In the absence of significant cooling, the minimum value of $\beta$ is likely to be achieved when ideal, monatomic gas pressure balances self-gravity in the transverse direction, in which case $\beta = 1/2$. Since compressive effects augment the density to the point where it is comparable to the initial density of the cloud when the COM recedes to the tidal radius (see Section \ref{sec:caustics}), the minimum rate at which the unbound filament density declines is 

\begin{equation}
\rho \simeq \rho_{\rm c}\left(t/T_{\rm tidal}\right)^{-3/2}, \quad T_{\rm tidal} \simeq \frac{r_{\rm t}^{3/2}}{\sqrt{GM_{\bullet}}},
\end{equation}
where $T_{\rm tidal}$ is on the order of the orbital time at the tidal radius (which, by construction, is comparable to the freefall time of the original cloud). The minimum time at which the unbound filament density drops to that of the (assumed-constant) background density, $\rho_{\rm a}$, is then

\begin{equation}
t_{\rm eq} \le T_{\rm tidal}\left(\frac{\rho_{\rm c}}{\rho_{\rm a}}\right)^{2/3}.
\end{equation}
This is a very rough approximation, which ignores, among other things, any radial variation in the background density profile. However, if the initial cloud density is, say, $10^3\times \rho_{\rm a}$, then we would expect the cloud to be able to last $\sim 100$ dynamical times before being sheared apart by the Kelvin-Helmholtz instability. For a cloud density of $10^{-20}$ g cm$^{-3}$, this timescale is $\sim 100$ Myr.

These arguments also assume that the filament is evolving only in the gravitational potential of the SMBH, but the steepening of the mass profile at distances $\gtrsim 60$ pc (see the discussion at the beginning of Section \ref{sec:model}) implies that the tidal field may become compressive, or at least very weakly disruptive, at distances roughly between $60 - 120$ pc. The compressive nature of the tidal field in this region inhibits the further lengthening of the filaments produced by the destruction of clouds by the disruptive potential near the GC. If the filament could survive the destabilizing effects of the shear present between it and the ambient gas, it could eventually decelerate to the point where it becomes approximately hydrostatic, which might suggest that filamentary structures produced in this way could be longer lived. However, hydrostatic, cylindrical structures are generically unstable under their own self-gravity\footnote{This is also true for any region of the filament that can remain self-gravitating. However, as described in \citet{coughlin20b}, the growth rate of the instability is inhibited by the shear along the filament axis, and the growth is no longer purely exponential.} in such a way that the growth rate of the instability is comparable to the freefall time across the filament width (e.g., \citealt{chandrasekhar53, ostriker64, breysse14, coughlin20b}). Thus, even in this scenario when the filament survives the destructive interactions with its environment and becomes hydrostatic, it likely fragments under its own self-gravity.

\subsection{Filament distribution and spatial orientation}
If the filamentary structures generated by the destruction of weakly bound (to the SMBH) clouds by the Galactic potential evolved in isolation and purely in the gravitational field of the SMBH, then the model presented here and the discussion in Section \ref{sec:lifetime} predicts that NTFs would eventually (i.e., when the cloud COM recedes well beyond the tidal radius) extend approximately \emph{radially} from the SMBH. This feature arises from the fact that, while the distances of individual fluid elements comprising the NTFs to the SMBH vary in such a way that they spread differentially over time (see Section \ref{sec:spread}), the conservation of angular momentum dictates that the spread in azimuthal angles of fluid elements around the marginally bound radius decreases in time. As such, the distribution of fluid elements becomes increasingly radially oriented as the filament expands; in the limit that the process responsible for feeding the clouds to the SMBH is isotropic, one would expect a distribution of filaments that points radially and isotropically away from the GC. The distribution of projected distances that we actually measure can then be calculated if the distribution of filament lengths is known. 

However, the distribution of NTFs and their spatial orientation is complicated by a number of effects. For one, the radial stretching only holds at sufficiently late times, while initially -- when the cloud COM is at a distance more comparable to the tidal radius -- the orientation is more random (see the bottom two panels of Figure 2 of \citealt{coughlin20} and Figure \ref{fig:cloud_disruption} here for a qualitative illustration). There is also a non-trivial dependence on the eccentricity of the COM, and while we have focused primarily on the case of marginally bound ($e = 1$) clouds, hyperbolic and eccentric initial cloud orbits result in lesser or greater degrees of curvature of the NTF, respectively. There is also the fact that the Galactic potential becomes dominated by additional, massive and stellar components as we recede to larger distances from the GC. Because the tidal force can become compressive when the mass profile is sufficiently steep, a filament can be torqued non-radially as the leading edge of the filament (i.e., at the largest distance) feels a \emph{larger} gravitational force than the trailing edge.

In addition, interactions between outgoing filaments and the matter in the GC will result in their deflection and reorientation, which was analyzed in detail in \citet{guillochon16} in the context of unbound debris streams generated from the tidal destruction of stars by a SMBH. For example, the encounter between the leading edge of the filament and an overdensity in the interstellar medium will preferentially decelerate that part of the filament, causing the NTF to bend in the vicinity of the overdensity. Indeed, this could be an explanation of the ``kink'' observed in the Snake and other filaments, specifically that such kinks arise when the filament interacts with a particularly dense region of ambient gas.

\subsection{Partial disruptions and the filament-cloud connection}
The previous sections considered the scenario in which the cloud is completely destroyed by the gravitational field (and the associated tides) of the GC. In this case, the entire cloud is transformed into a long, filamentary structure, with a spread of energies that depends on the orbital energy of the cloud COM.

The condition for a full disruption is that the pericenter distance of the cloud, $r_{\rm p}$, satisfies $r_{\rm p} \lesssim r_{\rm t}$, where $r_{\rm t}$ is the distance from the GC at which the differential gravitational force across the cloud diameter equals its own self-gravity (see Equation \ref{rtidal}). In general, the precise location at which the cloud is completely disrupted depends on its density profile, but simulations of stellar disruptions by SMBHs have found that $r_{\rm t}$ characterizes the full disruption distance to within a factor of $\sim 2$ depending on the properties of the star \citep{guillochon13, mainetti17}. On the other hand, if the pericenter distance of the star is larger than the radius at which it is completely destroyed, then a core either survives the entire encounter intact or reforms after the disruption, with the amount of material liberated from the surviving core a decreasing function of $\beta$ (see, e.g., Table 1 of \citealt{miles20}). 

If clouds can be completely destroyed by the GC gravitational potential, it seems unavoidable that partial disruptions of clouds should also occur at a rate that, while dependent on the mechanism responsible for feeding the clouds into the GC, is likely comparable to that at which full disruptions occur (see, e.g., the discussion in \citealt{coughlin19} for the case of tidal disruption events). In this case and in analogy with the stellar disruption scenario, one would expect two tidal tails of material to be shed from the cloud, with the mass contained in each tidal tail a function of the $\beta$ of the encounter. The overall evolution of the filamentary structures would be qualitatively similar to those from the full disruption case, but with the difference that the role of self-gravity will be reduced because of the smaller amount of mass and the greater amount of shear near the Hill sphere of the surviving core. 

The partial disruption of clouds provides an explanation for the observed association between some NTFs and molecular clouds (e.g., \citealt{bally87, morris96, yusef04, morris17}). Furthermore, the tidal compression of the cloud that occurs near pericenter may provide a perturbation that triggers star formation within the surviving cloud, creating a link between NTFs and star-forming regions. If the pericenter distance of the initial cloud is small, then the surviving cloud would be substantially perturbed, and the re-collapse of the cloud or re-accretion of the tidal material could lead to significant magnetic field amplification (e.g., \citealt{guillochon17, bonnerot17} for a discussion of this point in the context of TDEs). 

\section{Discussion and Conclusions}
\label{sec:summary}
We suggested that the numerous, nonthermal filamentary structures observed near the GC are produced by the destruction of molecular clouds by the gravitational potential associated with the GC. The main features of this model are summarized pictorially and qualitatively in Figure \ref{fig:cloud_disruption}. For simplicity, in much of the discussion of the model in Section \ref{sec:model} we assumed that the SMBH dominates the tidal field responsible for the destruction of the cloud. While this assumption is accurate for sufficiently small distances from the GC, at larger distances the stellar contribution to the enclosed mass dominates that of the SMBH; however, many of the bulk features and consequences of this model (e.g., the formation of a filamentary structure, the alignment of the magnetic field with the axis of the filament) are independent of this assumption.

In Section \ref{sec:model}, we broke the predictions of this model into various subsections in an attempt to highlight its salient and relevant features in the context of the observations of NTFs. Many of these features are, however, interrelated with one another and, in a more realistic model, must be considered simultaneously. For example, the width of the filament is constrained by the balance between pressure and self-gravity (or lack thereof), both of which depend on the density and the magnetic field, which themselves depend on the width and the rate at which the filament is stretched. When one contribution to the pressure dominates and the filament is self-gravitating, these effects can be isolated and approximate scaling solutions can be obtained for the evolution of, for example, the density as a function of time. Numerical simulations of the destruction of a magnetized cloud by the GC potential (or, more generally, a more detailed model) would more accurately determine the evolution of the system when these approximations cannot be made.

The discussion in Section \ref{sec:magnetic} describes the evolution of the magnetic field, and that because the filament is stretched preferentially in one direction, the magnetic field naturally aligns with its axis (see Figure \ref{fig:cloud_disruption}). Conditions are also likely to be conducive to magnetic reconnection and corresponding particle acceleration, which generates synchrotron emission. The strength of the magnetic field declines relatively slowly relative to the gas pressure, meaning the filament becomes magnetically dominated at late times, and the field strength is likely initially (i.e., when the filament recedes back out to a distance of $\sim r_{\rm t}$) comparable to the field of the original cloud (which arises from the fact that the gas is compressed within the orbital plane out to a distance of $\sim r_{\rm t}$; see the discussion in Section \ref{sec:caustics}). This result is consistent with claims made as early as \citet{roberts99}, who noted that the magnetic field strengths of NTFs made on dynamical grounds are comparable to the $\sim$ mG fields found in molecular clouds, which is also in line with more recent observations (e.g., \citealt{han17}). The linear morphology and relatively large strength of the magnetic fields in NTFs is therefore a natural consequence of this model, and has nothing to do with the putative existence of a large-scale, Galactic magnetic field.

An additional property of a fraction of filaments is that they show some degree of multiplicity or substructure, with more than one filament closely separated in space. As pointed out by \citet{yusef04}, one possible explanation for this occurrence is that there is actually only one filament, but edge-brightened emission results in two seemingly distinct structures. Another possibility is that two filaments may appear to be neighboring in their projected positions, but their physical separation is much larger. 

While these explanations hint at the notion that the substructure and multiplicity of filaments is only apparent, a number of closely spaced filaments could arise from substructure within the original cloud. Specifically, as noted in Section \ref{sec:model}, Equation \eqref{rtidal} gives the tidal disruption radius of a cloud with \emph{average} density $\rho_{\rm c}$, but multiple over-densities could exist within the cloud, each with a tidal radius smaller than that of the cloud on average. In this case, one would expect each high-density pocket to remain relatively intact until it reaches its own tidal radius, which would eventually occur for a sufficiently large value of $\beta$ of the cloud COM. In this case and with a range of over-densities each with its own tidal disruption radius, each pocket of dense material would be disrupted in relatively close temporal succession, resulting in the formation of multiple filaments with similar spatial orientation and overall extent. 

In our model we treated the cloud as a spherical object, whereas realistic molecular clouds are observed to possess a variety of aspherical deformations. We also treated the potential of the GC as spherical, while the flattening of the galactic disc implies that this approximation cannot be upheld exactly (e.g., \citealt{launhardt02, sormani20}). These additional features (and many others) will further complicate the relatively simple picture outlined here, and future investigations should address the impact of these complications on the properties of the radio filaments generated by this process.

\section*{Acknowledgements}
We thank Jim Pringle for useful comments on an early draft, and the anonymous referee for useful suggestions that helped clarify the paper. E.R.C.~acknowledges support from the National Science Foundation through grant AST-2006684. C.J.N. is supported by the Science and Technology Facilities Council (grant number ST/M005917/1). This project has received funding from the European Union’s Horizon 2020 research and innovation program under the Marie Sk\l{}odowska-Curie grant agreement No 823823 (Dustbusters RISE project). This material is based in part upon work supported by the National Science Foundation under Grant No. 2008101 to A.~G.

\section*{Data Availability}
No new data were generated or analysed in support of this research.

\bibliographystyle{mnras}

\bsp	
\label{lastpage}
\end{document}